\documentclass[twocolumn]{aastex631}

\usepackage{amsmath}
\begin{document}

\title{\texttt{FREmu}: Power Spectrum Emulator for $f(R)$ Gravity}

\author{Jiachen Bai}
\affiliation{Department of Astronomy, Beijing Normal University, Beijing 100875, China}

\author{Junqing Xia}
\affiliation{Department of Astronomy, Beijing Normal University, Beijing 100875, China}
\affiliation{Institute for Frontiers in Astronomy and Astrophysics, Beijing Normal University, Beijing 100875, China}



\begin{abstract}
To investigate gravity in the non-linear regime of cosmic structure using measurements from Stage-IV surveys, it is imperative to accurately compute large-scale structure observables, such as non-linear matter power spectra, for gravity models that extend beyond general relativity. However, the theoretical predictions of non-linear observables are typically derived from N-body simulations, which demand substantial computational resources. In this study, we introduce a novel public emulator, termed \texttt{FREmu}, designed to provide rapid and precise forecasts of non-linear power spectra specifically for the Hu-Sawicki $f(R)$ gravity model across scales $0.0089 h \mathrm{Mpc}^{-1}<k<0.5 h \mathrm{Mpc}^{-1}$ and redshifts $0<z<3$. \texttt{FREmu} leverages Principal Component Analysis and Artificial Neural Networks to establish a mapping from parameters to power spectra, utilizing training data derived from the Quijote-MG simulation suite. With a parameter space encompassing 7 dimensions, including $\Omega_m$, $\Omega_b$, $h$, $n_s$, $\sigma_8$, $M_{\nu}$ and $f_{R_0}$, the emulator achieves an accuracy exceeding 95\% for the majority of cases, thus proving to be highly efficient for constraining parameters.
\end{abstract}


\section{Introduction}

The discovery of late-time acceleration of the universe \citep{1998AJ....116.1009R} has presented a significant challenge to contemporary cosmology for over two decades. Tensions such as the Hubble tension and the $\sigma_8$ tension  \citep{2020A&A...641A...6P} have also cast doubts on the standard $\Lambda$CDM model of cosmology. In response to these challenges, alternative models beyond the $\Lambda$CDM model have been postulated to address these issues. Numerous dark energy (DE) models have been proposed to tackle these problems, and modifications to the theory of gravity are also considered as potential avenues to unlock the mysteries of the universe.

Modified gravity (MG), as a departure from the general theory of relativity (GR), has been under consideration for decades to resolve tensions in cosmology. Its cosmological implications can be tested through observations of the background evolution and the large-scale structure (LSS) of the universe. Given that gravity plays a pivotal role in structure formation, the gravitational effects of MG can manifest observables that diverge from those predicted by GR. Notably, MG alters the Poisson equations, leading to variations in the gravitational potential derived from GR, consequently influencing the evolution of LSS. Moreover, certain MG theories give rise to screening mechanisms at relatively small scales, such as the Vainshtein screening from the Dvali-Gabadadze-Porrati (DGP) theory \citep{2000PhLB..485..208D,2013CQGra..30r4001B} and chameleon screening from the $f(R)$ theory \citep{1970MNRAS.150....1B,2004PhRvD..69d4026K,2018LRR....21....1B} which introduce differences in LSS formation at smaller scales.

Massive neutrinos, fundamental constituents of the universe, also influence the formation and evolution of large-scale structures. Due to their non-zero masses, massive neutrinos exhibit behavior similar to cold dark matter (CDM) on large scales. However, their free-streaming nature suppresses structure formation on smaller scales relative to CDM \citep{2010PrPNP..64..360F}. Despite being less dense than other forms of matter, massive neutrinos still give rise to small-scale overdensities, and their clustering effects have been discerned through N-body simulations \citep{1982PhRvL..48..894M}. Within the framework of $f(R)$ gravity discussed herein, the effects of massive neutrinos and MG on LSS formation exhibit strong degeneracy in simulation outcomes \citep{2014MNRAS.440...75B}, complicating the cosmological constraints on these parameters.

Traditionally, observations of LSS primarily focused on large scales, offering only rudimentary constraints on cosmological models. The dearth of information at small scales, where non-linear effects are prominent, significantly limited the constraints on MG models. This limitation arises because certain MG models exhibit substantial deviations from GR at small scales, and the effects of massive neutrinos are predominantly manifested at small scales. Presently, observations of LSS are entering a new epoch with the advent of stage-IV surveys like DESI \citep{2016arXiv161100036D}, LSST \citep{2019ApJ...873..111I}, CSST \citep{2019ApJ...883..203G}, and Euclid \citep{2011arXiv1110.3193L}, anticipated to furnish precise observables on small scales of LSS with expanded survey volume, deeper redshift, greater magnitude depth, and increased observational sources. These enhanced observations are poised to make substantial contributions to model constraints, particularly concerning MG theories. To effectively leverage these experiments, accurate theoretical predictions of observables with non-linear effects, such as non-linear matter power spectra, are imperative for conducting parameter constraints using statistical techniques like Markov chain Monte Carlo (MCMC) methods \citep{1970Bimka..57...97H}. In contrast to linear regimes, power spectra in non-linear regimes are not amenable to analytical solutions and are commonly derived through cosmological N-body simulations. These simulations necessitate significant computational resources, as their precision hinges on the temporal discretization and spatial resolution employed. Thus, there arises a pressing need to devise an alternative methodology that can efficiently and accurately forecast observables, such as matter power spectra, to effectively constrain MG models.

To meet this requirement, emulators have been introduced \citep{PhysRevD.76.083503} to provide direct predictions from parameter space without necessitating N-body simulations. Over the years, various emulators have been developed to forecast matter power spectra for diverse models, including \texttt{PkANN} \citep{2014MNRAS.439.2102A}, \texttt{EuclidEmulator} \citep{2019MNRAS.484.5509E}, and \texttt{FORGE} \citep{2022MNRAS.515.4161A} and \texttt{E-MANTIS} \citep{2024MNRAS.527.7242S} for $f(R)$ gravity, as well as \texttt{nDGPemu} \citep{2023JCAP...12..045F} for DGP gravity.

In this study, our focus centers on the specific $f(R)$ model introduced by \citet{2007PhRvD..76f4004H} (HS), where the deviation from GR is encapsulated by a single parameter, $f_{R_0}$. We have developed a new public emulator, dubbed \texttt{FREmu}, to deliver swift and accurate predictions of matter power spectra for the HS model incorporating massive neutrinos. This emulator is founded on simulation data, specifically the Quijote-MG simulation suite \citep[in prep]{baldi2024}, which encompasses 2048 sets of simulations sampled with Sobol sequences across a wide parameter space and executed using the MG-Gadget N-body simulation code \citep{2013MNRAS.436..348P}. To efficiently predict the simulated power spectra from parameters, we initially convert them to power spectrum boosts relative to those computed with the halofit method \citep{2021MNRAS.502.1401M} for $\Lambda$CDM cosmologies. Subsequently, we employ Principal Component Analysis (PCA) to reduce the dimension of the target space and harness Artificial Neural Networks (ANNs) to evaluate the coefficients of each principal component for any given parameters. By combining these components, we derive boosts, multiply them with halofit power spectra, and ultimately provide predictions of non-linear matter power spectra for the HS $f(R)$ model, maintaining relative errors below 5\% under most conditions. Notably, our emulator encompasses a significantly broader parameter space, both in terms of dimensionality (encompassing 7 parameters like $\Omega_m$, $\Omega_b$, $h$, $n_s$, $\sigma_8$, $M_{\nu}$, and $f_{R_0}$) and ranges, compared to previously mentioned emulators. 

The paper is organized as follows: Section \ref{sec:fr} delves into the theoretical underpinnings of the $f(R)$ model, elucidating its implications on structure formations and its current constraints to underscore the necessity of our work. Section~\ref{sec:methods}, outlines the design of experiments (DOE) of Quijote-MG simulations and the methodologies employed in constructing our emulator. Subsequently, in Section~\ref{sec:results}, we demonstrate the accuracy of our emulator and present an application example showcasing its prowess in constraining parameters. Finally, in Section~\ref{sec:conclusions}, we conclude our work and outline potential avenues for future developments.

\section{The Hu \& Sawicki f(R) gravity}\label{sec:fr}

In $f(R)$ gravity \citep{1970MNRAS.150....1B}, which stands as an alternative theory to General Relativity (GR), gravitational dynamics are delineated not solely by the Ricci scalar but also by an additional modification term reliant on the Ricci scalar. This augmentation injects supplementary degrees of freedom into the framework, engendering disparities from the forecasts of GR, notably evident on cosmological magnitudes. To gain a concise comprehension of this theory, the mathematical formalism and physical ramifications of $f(R)$ gravity are expounded upon in this section.

The fundamental principle commonly employed to elucidate natural phenomena is the principle of least action, which also governs gravity. In $f(R)$ gravity, this principle is encapsulated by the Einstein-Hilbert action augmented with an extra $f(R)$ term, expressed as:
\begin{equation}
S=\int \mathrm{d}^4 x \sqrt{-g}\left[\frac{R+f(R)}{16 \pi G}+\mathcal{L}_{\mathrm{m}}\right]
\end{equation}
Here, $g$ signifies the determinant of the metric tensor $g_{\mu\nu}$, $R$ represents the Ricci scalar, $f(R)$ denotes a function of the Ricci scalar, $G$ stands for the gravitational constant, and $\mathcal{L}_{\mathrm{m}}$ denotes the Lagrangian density of matter fields.

The field equations are derived by varying the action with respect to the metric tensor, resulting in modified gravitational field equations that couple the Einstein tensor $G_{\mu\nu}$ with terms involving derivatives of $f(R)$ with respect to $R$. These equations are represented as:
\begin{equation}
G_{\mu\nu} +f_{R}R_{\mu\nu} - \left(\frac{f}{2}-\Box{f_{R}}\right)g_{\mu\nu}-\nabla_{\mu}\nabla_{\nu}f_{R} = \kappa^{2} T_{\mu\nu}
\end{equation}
Here, $R_{\mu \nu}$ denotes the Ricci tensor, $G_{\mu \nu}$ represents the Einstein tensor, $T_{\mu \nu}$ signifies the energy-momentum tensor of matter, $\nabla_\mu$ denotes the covariant derivative compatible with the metric, $\square \equiv g^{\mu \nu} \nabla_\mu \nabla_\nu$, and $f_R$ is the derivative of $f(R)$ with respect to $R$, referred to as the scalar degree of freedom (SDOF).

In the quasi-static regime, these modified field equations can be approximated to yield the modified Poisson equation and the equation of motion for the SDOF:
\begin{align}
    &\boldsymbol{\nabla}^2 \Phi=\frac{16 \pi G}{3} \delta \rho-\frac{1}{6} \delta R\\
    &\boldsymbol{\nabla}^2 f_R=\frac{1}{3}(\delta R-8 \pi G \delta \rho)
\end{align}
where $\Phi$ denotes the Newtonian potential, $\delta \rho$ and $\delta R$ represent perturbations to the matter density and Ricci scalar respectively, and $\boldsymbol{\nabla}$ is the three-dimensional gradient operator.

To parameterize the function $f(R)$, a commonly used form, proposed by \citet{2007PhRvD..76f4004H}, is
\begin{equation}
    f(R)=-m^2 \frac{c_1\left(R / m^2\right)^n}{c_2\left(R / m^2\right)^n+1}
\end{equation}
where $c_1$, $c_2$, and $n$ are dimensionless parameters, and $m$ is a curvature scale defined as
\begin{equation}
    m^2 \equiv \frac{\Omega_m H_0^2}{c^2}
\end{equation}
here, $\Omega_m$ represents the current fractional matter density, and $H_0$ is the Hubble constant. By appropriately choosing the parameters $c_1$ and $c_2$, one can ensure that the background evolution mimics that of the $\Lambda$CDM model.

In such scenarios, the SDOF can be approximated as
\begin{equation}
    f_R=-n \frac{c_1}{c_2^2}\left(\frac{m^2}{R}\right)^{n+1}
\end{equation}
where $n = 1$ in our analysis. The parameter ${c_1}/{c_2^2}$, often denoted as $f_{R_0}$, represents the current background value of the SDOF.

This parameterization of $f(R)$ gravity enables a detailed exploration of its impacts on various astrophysical and cosmological phenomena, shedding light on gravity's behavior beyond General Relativity and its consequences for the evolution and formation of cosmic structures.

In ongoing cosmological studies, the constraints on HS $f(R)$ gravity theory have certain limitations,  with the current upper bound on $\lvert f_{R_0} \rvert$ found to be $\lvert f_{R_0} \rvert  \leq 5.68 \times 10^{-7}$ \citep{2021PhRvD.104j3519L}.

While past LSS data offer crucial insights into cosmic evolution, they may not robustly delimit the small-scale effects within the HS $f(R)$ theory. These effects, such as chameleon screening \citep{2004PhRvD..69d4026K}, could notably impact the behaviors of LSS evolution at small scales. Future Stage-IV surveys are anticipated to furnish more precise data at smaller scales through LSS observations. Predicting observables at small scales, encompassing non-linear effects like non-linear power spectra, could significantly bolster the constraints on the $f(R)$ model \citep{2023arXiv230611053C}.

\section{Methods}\label{sec:methods}

\subsection{The Quijote-MG simulations}
While it is challenging to analytically solve the non-linear matter power spectra in $f(R)$ gravity, N-body simulations offer a viable alternative with high resolutions that can accommodate large overdensities beyond linear limitations. However, conducting simulations is a resource-intensive task, making it impractical for everyone to run simulations and inefficient to perform parameter constraints that involve sampling numerous parameter combinations.

The concept behind our emulator involves constructing a rapid model to forecast non-linear power spectra by leveraging labeled data from existing simulation results. This approach utilizes machine learning algorithms to uncover the relationships between cosmological parameters and the non-linear matter power spectra of LSS. By doing so, we aim to streamline the process and reduce the computational burden associated with parameter constraints in $f(R)$ gravity studies.

The emulator here is built upon the Quijote-MG simulation suite developed by \citet[in prep]{baldi2024}. Quijote-MG comprises 2,048 N-body simulations executed with the MG-Gadget code \citep{2013MNRAS.436..348P}, utilizing HS $f(R)$ gravity as the MG model. Each simulation tracks the evolution of $512^3$ dark matter particles alongside $512^3$ neutrinos within a periodic cosmological volume of $V = (1000 \mathrm{Mpc} / h)^3$. The initial conditions are generated at redshift $z=127$ using the Zel'dovich approximation \citep{1970A&A.....5...84Z}, and the simulations incorporate the appropriate Hubble function $H(z)$. These simulations store 5 snapshots at redshifts $0, 0.5, 1, 2$, and $3$, providing valuable data for training and validating the emulator to predict non-linear matter power spectra in the context of HS $f(R)$ gravity.

In each MG simulation within the Quijote-MG suite, a distinctive set of initial random seed values is assigned along with parameters that define the cosmological model. These parameters encompass the matter density parameter ($\Omega_{\mathrm{m}}$), baryon density parameter ($\Omega_{\mathrm{b}}$), Hubble parameter ($h$), spectral index of primordial density fluctuations ($n_s$), the standard deviation of density perturbations within an 8 ${\rm Mpc}/h$ sphere ($\sigma_8$), the sum of neutrino masses ($M_\nu$), and the present-day background value of the scalar degree of freedom ($f_{R_0}$).

To thoroughly explore the parameter space, Quijote-MG employs a Sobol sequence \citep{sobol1967distribution}, which is a low-discrepancy sequence ensuring more uniform coverage of the parameter space compared to random sampling. This systematic sampling method facilitates an effective exploration of the parameter space while ensuring uniformity and preventing point clustering.

\begin{figure*}
    \centering
	\includegraphics[width=1.7\columnwidth]{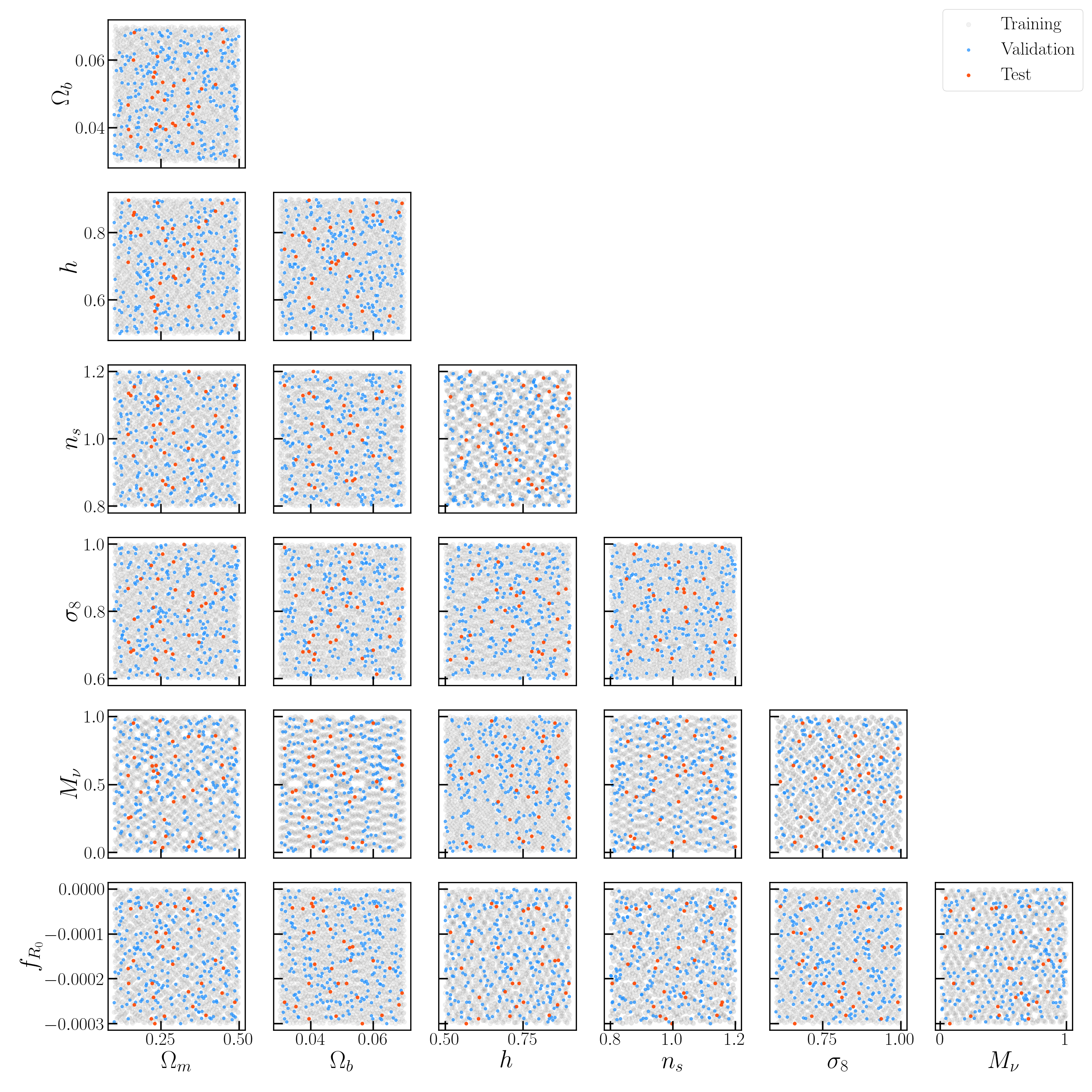}
    \caption{ Parameter space of the Quijote-MG simulation suite. The grey dots are 1760 samples of the training set, the blue dots are 256 samples of the validation set, and the red dots are 32 samples of the test set. All 3 parameter sets cover the entire parameter space well.}
    \label{fig:samples}
\end{figure*}

As depicted in Fig.\ref{fig:samples}, the parameter values in the simulations are constrained within specific ranges to cover wide priors without influencing subsequent inference. The ranges of parameters are defined as reasonable bounds for the input of our emulator:
\begin{equation}
	\begin{aligned}
		0.1 & \leq \Omega_{\mathrm{m}} \leq 0.5\\
		0.03 & \leq \Omega_{\mathrm{b}} \leq 0.07\\
		0.5 & \leq h \leq 0.9\\
		0.8 & \leq n_s \leq 1.2\\
		0.6 & \leq \sigma_8 \leq 1.0\\
		0.01 & \leq M_\nu[\mathrm{eV}] \leq 1.0 \\
		-3 \times 10^{-4} & \leq f_{R_0} \leq 0
	\end{aligned}
\end{equation}

The systematic DOE approach employed ensures that our emulator can accurately predict matter power spectra for any parameter set within the specified ranges, maintaining consistent accuracy across the large parameter space. This organized sampling strategy enhances the efficiency of emulating in such a vast parameter domain.

The Quijote-MG simulations yield matter power spectra for each parameter set at 5 redshift nodes ($z=0.0, 0.5, 1.0, 2.0, 3.0$) spanning from the largest to the smallest scales in the simulation box ($0.0089 h \mathrm{Mpc}^{-1} < k < 1.58 h \mathrm{Mpc}^{-1}$). The emulator provides predictions within these corresponding ranges. Notably, the minimum trustworthy scale is $k=0.5 h \mathrm{Mpc}^{-1}$, constrained by the current resolution of the simulations. However, for demonstration purposes, we consider $k$ values exceeding this limit to showcase the potential of emulator to offer power spectra at smaller scales with enhanced resolution. Subsequently, we will continue the discussion with a relatively large limit for $k$.

\subsection{Emulation}
\subsubsection{MG power spectrum boosts}
\begin{figure}
    \centering
	\includegraphics[width=1.\columnwidth]{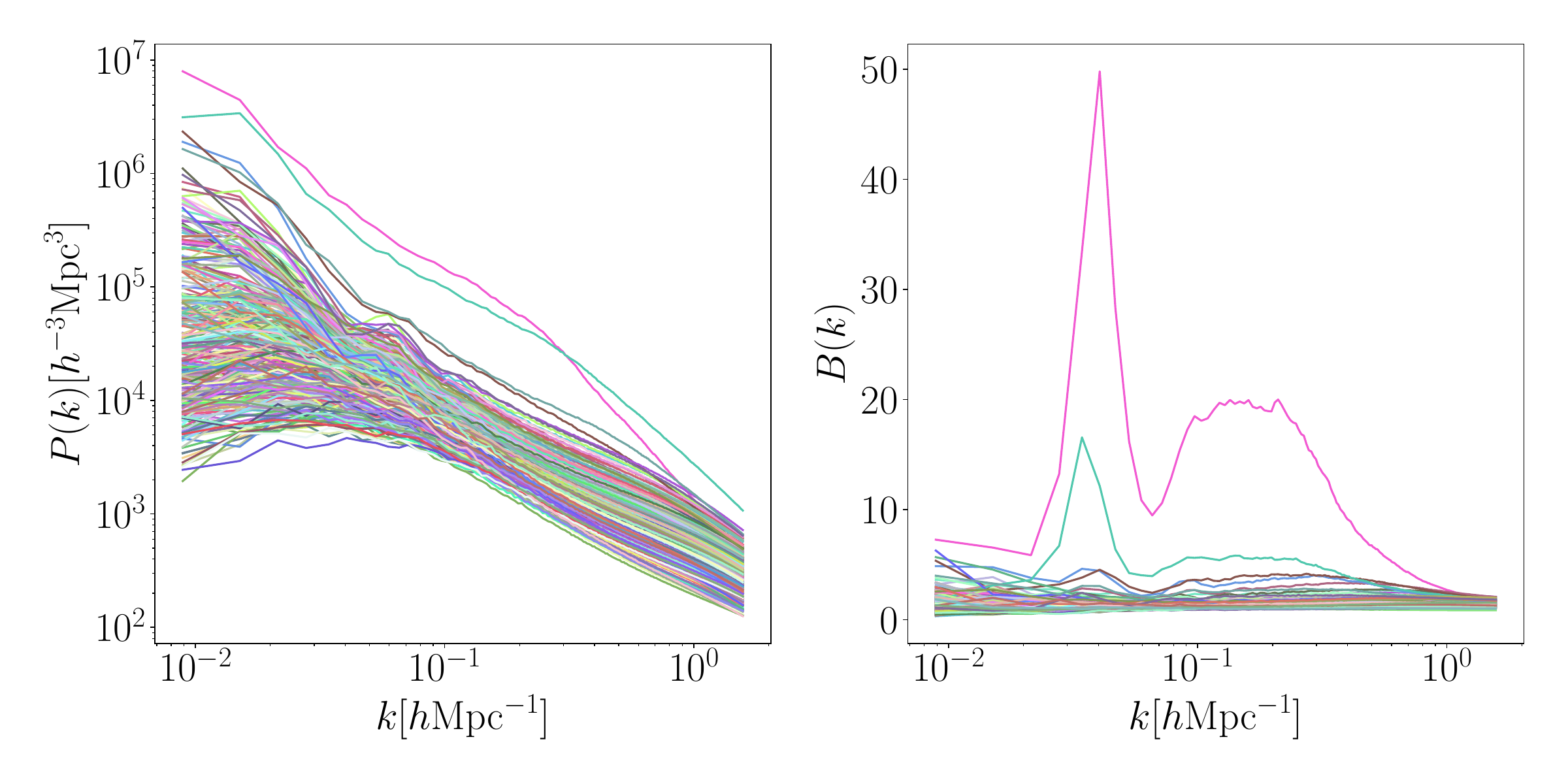}
    \caption{Comparison between the original power spectra (left figure) ranging from $10^2$ to $10^7$ and the power spectrum boosts (right figure) ranging from 0 to 50. Different colors represent $P(k)$ and $B(k)$ of different samples, respectively.}
    \label{fig:pkbk}
\end{figure}

The extensive range of emulation required can be estimated based on the data provided, as illustrated in the left panel of Figure~\ref{fig:pkbk}. For clarity and ease of discussion, all results presented here are at redshift $z = 0$. The broad target range poses a significant challenge to our emulation process, given that power spectra can vary by up to 4 orders of magnitude even at the same scale. This variability makes it challenging to establish precise correspondences between the parameter space and power spectra space.

To address this challenge, we adopt a two-step approach leveraging the halofit method outlined by \citet{2021MNRAS.502.1401M} for the $\Lambda$CDM model. Rather than directly emulating the power spectra, we focus on estimating the MG power spectrum boosts, denoted as $B(k)$, defined as the ratio of the non-linear power spectra of $f(R)$ models to the halofit-calculated non-linear power spectra of $\Lambda$CDM models:
\begin{equation}
B(k) = \frac{P^{\mathrm{nonlin}}_{f(R)+M_{\nu}}(k)}{P^{\mathrm{halofit}}_{\Lambda \mathrm{CDM}}(k)}
\end{equation}
Here, $P^{\mathrm{nonlin}}_{f(R)+M_{\nu}}(k)$ represents the non-linear power spectra of $f(R)$ models with massive neutrinos, while $P^{\mathrm{halofit}}_{\Lambda \mathrm{CDM}}(k)$ signifies the halofit-calculated results for non-linear power spectra in $\Lambda$CDM models with massless neutrinos.

The halofit model, a semi-analytical approach based on the halo model, is commonly used to estimate the non-linear matter power spectra in $\Lambda$CDM cosmologies. By leveraging halofit results as a benchmark, we can assess the impacts of $f(R)$ gravity and neutrino mass on the nonlinear power spectrum. These effects, illustrated in the narrower ranges shown in the right panel of Figure~\ref{fig:pkbk}, are more predictable and manageable. While some emulators, like \texttt{EuclidEmulator} \citep{2019MNRAS.484.5509E}, utilize linear results as references for boosts, our scenario involves samples with significant deviations in power spectrum values due to non-linear, MG, or massive neutrino effects. This can lead to wide-ranging boosts that complicate predictions and reduce accuracy. As discussed in Section~\ref{sec:compare}, comparing methods reveals that using halofit has significantly enhanced the accuracy of our results.

For our technical implementation, we rely on the publicly available code \texttt{CAMB} \citep{2011ascl.soft02026L}, which features built-in halofit modules.  Specifically, we employ the \texttt{mead2020} module, to directly obtain the reference power spectra required.

This targeted approach optimizes computational resources towards specific features of interest, enabling more precise and efficient emulation of $f(R)$ gravity and massive neutrino effects within the cosmological framework. Moreover, it enhances the ability of machine learning algorithms to analyze the impacts of MG and massive neutrinos on matter power spectra.

Following the emulation of the boost factor for a parameter sample in the initial step, we can then combine it with the halofit result to derive the non-linear matter power spectrum incorporating $f(R)$ and massive neutrino effects, representing the output of our emulator. This two-step emulation strategy enhances the accuracy and reliability of our predictions.

\subsubsection{Principal Component Analysis (PCA)}
Upon converting power spectra to power spectrum boosts, our objective is to establish a mapping from a 7-dimensional parameter space to a boost space spanning hundreds of dimensions. To mitigate computational complexity, Principal Component Analysis (PCA) has been integrated into our methodology to tackle the issue of high dimensionality within the target space. The fundamental concept involves decomposing the boost data into a collection of predetermined orthonormal basis functions alongside their corresponding coefficients (see \citet{jolliffe2016principal} for more details). This decomposition enables us to express the power spectrum boosts as a linear amalgamation of these basis functions with minimal error. By keeping the basis functions unchanged, our focus shifts to comprehending the dynamics of coefficient interdependence on parameters, significantly alleviating the complexity associated with prediction tasks.

In mathematical terms, the power spectrum boost $B(k, z ; \theta)$ is formulated as follows:
\begin{equation}
B(k, z ; \theta) = \mu_{B}(k, z) + \sum_{i=1}^{N_{\mathrm{pc}}} \phi_i(k, z) w_i(\theta) + \epsilon
\end{equation}
In this expression, $\mu_{B}(k, z)$ denotes the mean value of all boost data, $\phi_i(k, z)$ represents the orthonormal basis functions, $w_i(\theta)$ corresponds to the coefficients associated with each basis function, and $\epsilon$ factors in the decomposition error. The selection of the number of basis functions, denoted as $N_{\mathrm{pc}}$, is determined to be smaller than the number of scale nodes $N_k$ to enhance computational efficiency.

In our methodology, we have chosen to incorporate 24 principal components (PCs) in conjunction with the mean function to effectively capture the variance within the dataset. This strategic choice serves to mitigate the complexity of emulation, requiring the prediction of only 24 coefficients for each sample with 7 parameters. By reducing the dimensionality of the target space while preserving crucial information through PCA, we optimize the emulation process and enhance predictive efficiency. To elucidate this improvement, Section~\ref{sec:compare} offers a comparative analysis highlighting the accuracy variances observed when employing PCA versus when it is omitted. Through the application of PCA, we observe consistent accuracy levels alongside improved operational efficiency.

\subsubsection{Artificial Neural Network (ANN)}
In the context of predicting PCA coefficients from parameters, we have employed Artificial Neural Networks (ANNs) to interpolate PCA coefficients based on simulation data.

Artificial Neural Networks are computational models inspired by the neural networks in the human brain, comprising interconnected nodes called neurons arranged in layers, including input, hidden, and output layers. Information flows from the input layer through the hidden layers to the output layer.

Each neuron receives input signals, applies weights to these signals, sums them, and applies an activation function to generate an output. This output is then passed to the subsequent layer as input. By adjusting the weights of connections between neurons, ANNs can learn to map input data to desired outputs, making them invaluable for tasks like classification, regression, and pattern recognition.

Training an ANN involves optimizing its weights to minimize a predefined objective function, often termed the loss function, typically achieved through backpropagation and gradient descent or its variants.

In our study, we employ a neural network with a single hidden layer (referred to as a shallow neural network), featuring a linear input layer with 7 neurons (representing 7 cosmological parameters) and a linear output layer with 24 neurons (representing 24 PCA coefficients). We use the sigmoid function as the activation function in the hidden layer, the size of which is determined by our hyperparameter study discussed later. Additionally, we utilize the Adam optimizer (see \citet{2014arXiv1412.6980K} for more details), the Mean Squared Error (MSE) loss function, and the StepLR scheduler to adjust the learning rate during training.

To optimize hyperparameters and identify the best architecture, we partition the dataset of 2048 samples into  1760 training samples, 256 validation samples, and 32 test samples. The training sets are used for weight optimization, the validation sets assess model performance with varied hyperparameter combinations, and the test sets evaluate the emulator's accuracy on new samples. As depicted in Fig.\ref{fig:samples}, all 3 sets of samples are uniformly distributed, making our training results reliable across all parameter ranges.

Hyperparameter optimization involves adjusting the hidden layer size ($N_{\mathrm{hidd}}$), the learning rate (Lr), the reduction percentage of learning rate every 1000 epochs ($\gamma$), and the total training epochs within the hyperparameter space. We randomly sample parameters to determine optimal values with the least MSE loss on validation sets. Our optimization process leverages the Python libraries \texttt{PyTorch} \citep{2019arXiv191201703P} and \texttt{optuna} \citep{akiba2019optuna}. For our study, we set $N_{\mathrm{hidd}}=1020$, ${\rm Lr}=0.04$, $\gamma=0.8$, and ${\rm epochs}=20000$ based on our optimization results.

\subsubsection{Emulation for arbitrary redshifts}
For the analysis of redshift nodes, we conducted individual training sessions for our ANN using consistent hyperparameters to generate forecasts at redshift values of $z=0.0, 0.5, 1.0, 2.0, 3.0$. Nonetheless, limiting our predictions to only five specific redshift nodes proves inadequate for our practical requirements. It is imperative that our emulator possesses the capability to interpolate predictions between these predefined redshift nodes.

In order to expand the emulation scope beyond the established redshift nodes, we have incorporated a cubic interpolation methodology. Cubic interpolation represents a numerical algorithm employed to approximate values between known data points. This technique involves the fitting of a cubic polynomial function to a subset of neighboring data points, which is subsequently utilized to interpolate values at desired positions within the data range.

To elaborate further, when considering an arbitrary redshift, we initially calculate the power spectrum boosts for the five nodes that have been trained with data. Subsequently, we utilize these five nodes to interpolate and determine the boost at the desired redshift. Once the boost value is obtained, it is multiplied with the halofit result computed using the \texttt{CAMB} software at the specified redshift. This process allows for the derivation of the non-linear matter power spectrum at the arbitrary redshift.

The efficacy of such an interpolation technique is elucidated in section~\ref{sec:interp}. This approach to interpolation enables accurate estimation of the power spectrum boost at intermediary redshifts by leveraging the information embedded within the trained ANN models.

With the integration of interpolation, our emulator is capable of furnishing non-linear matter power spectra within a redshift spectrum ranging from $z=0.0$ to $z=3.0$. This enhancement renders the emulator exceptionally practical, facilitating the connection between theoretical projections and observational data, thereby facilitating parameter constraints.

\section{Results}\label{sec:results}
\subsection{Accuracy test}
Upon training our emulator with 1760 training sets, predictions of power spectra can be made in about 1 second for each given set of parameters. This kind of efficiency greatly exceeds that of simulations, whether conventional N-body simulations (about $10^3$ hours) or COLA simulations \citep{2017JCAP...08..006W} (about 1 hour). In this section, we evaluate our emulator's performance on 32 test samples to validate its capacity for rendering precise predictions for new parameter samples.

\begin{figure*}
    \centering
	\includegraphics[width=1.7\columnwidth]{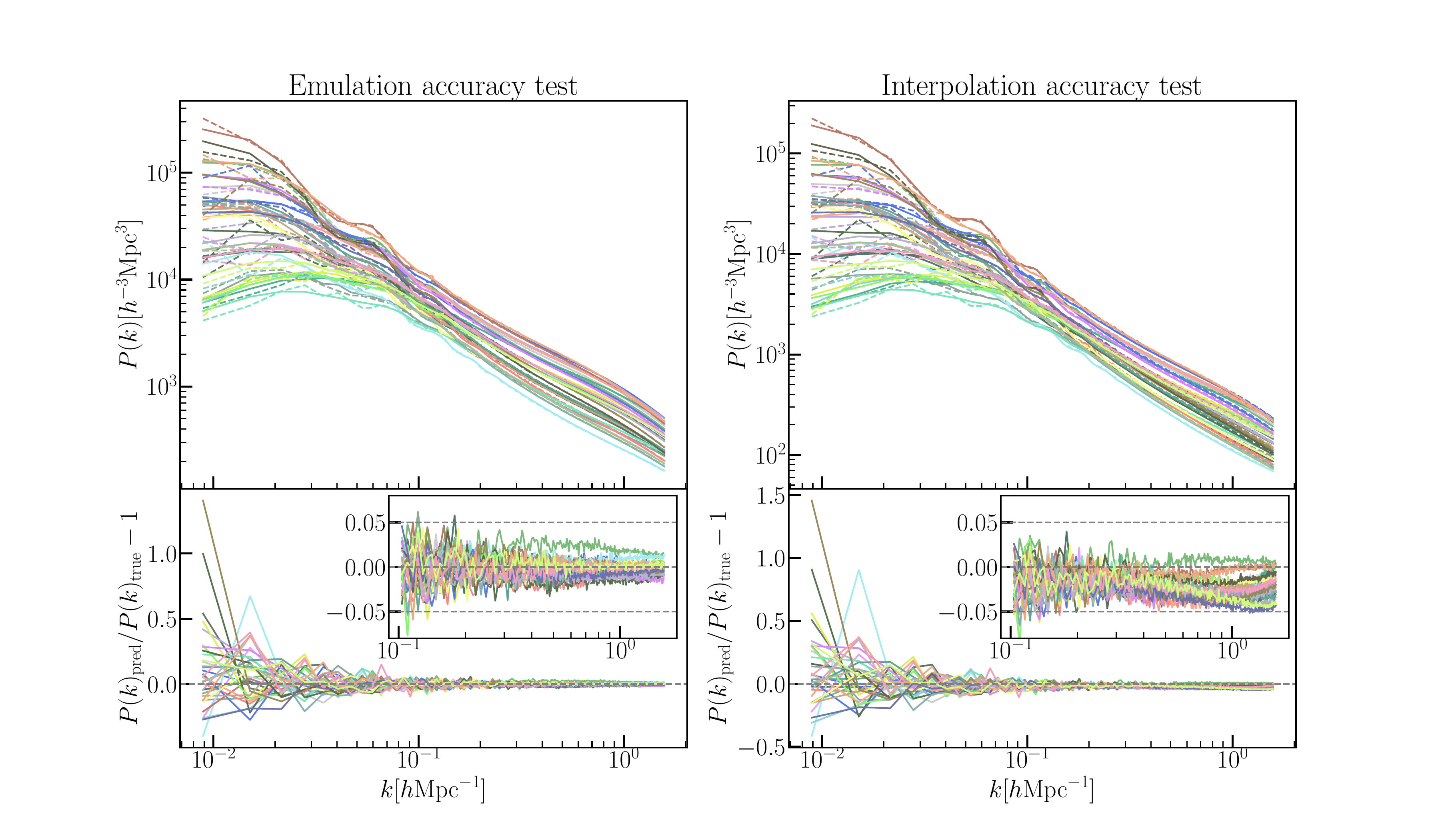}
    \caption{Visual representations of emulation accuracy (on the left) and interpolation accuracy (on the right) using 32 test samples} are presented. The dashed lines represent the true values of power spectra, whereas the solid lines, color-coded accordingly, depict the predicted values. The corresponding relative errors are displayed below, with particular emphasis on errors at small scales in a sub-figure, where 5\% errors are delineated by grey dashed lines.
    \label{fig:combine_1}
\end{figure*}

The precision assessments, illustrated in the left panel of Figure~\ref{fig:combine_1}, elucidate the efficacy of our emulator. The dashed lines delineate the true power spectrum values, whereas the solid lines, color-coded correspondingly, represent the predictions generated by our emulator. Furthermore, the respective relative errors are delineated below. Notably, at large scales where $k < 10^{-1} h \, \text{Mpc}^{-1}$, our emulator demonstrates errors exceeding 5\%, primarily attributed to significant cosmic variances. These variances arise from the restricted number of samples available for simulating initial conditions. Specifically, given the presence of only one simulation per parameter sample, considerable cosmic variances persist due to insufficient sample sizes to mitigate them.

Conversely, at smaller scales where $k > 10^{-1} h \, \text{Mpc}^{-1}$, the relative errors of our emulator predominantly remain below 5\% or even lower. This observation underscores the exceptional performance of our emulator, particularly in accurately predicting power spectra at smaller scales.

This analysis unveils the emulator's proficiency in accurately replicating power spectra across diverse scales and cosmological parameters. Although slight discrepancies may arise, especially at larger scales or under specific cosmological scenarios, the emulator consistently showcases exceptional precision in capturing the intrinsic patterns and characteristics of the power spectra.

\begin{figure}
    \centering
	\includegraphics[width=1.\columnwidth]{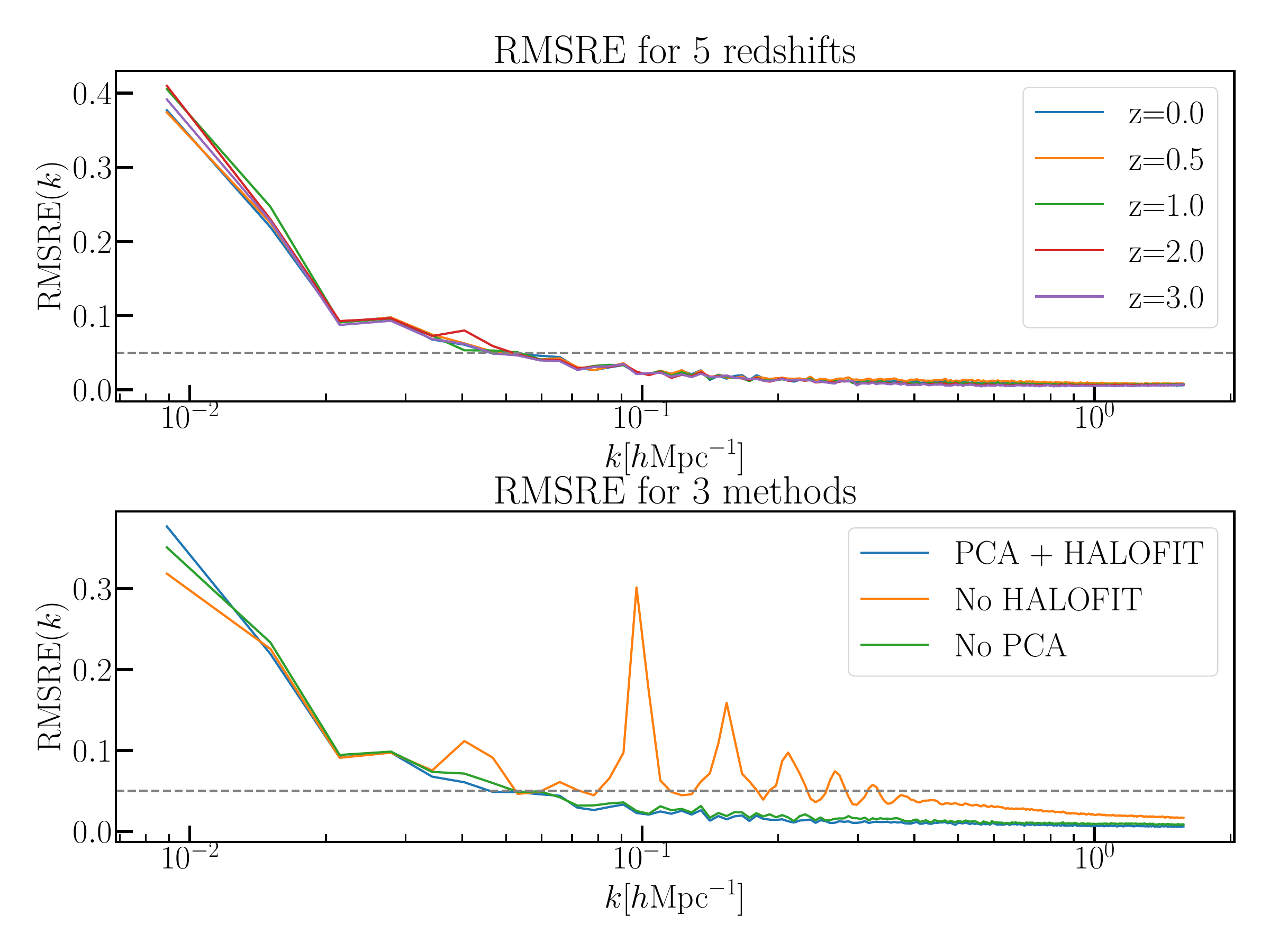}
    \caption{The upper figure illustrates the RMSREs of the emulator at 5 redshifts, highlighting 5\% errors through dashed lines. In the lower figure, the accuracy of various emulation strategies is compared, also featuring 5\% errors delineated for reference.}
    \label{fig:combine_2}
\end{figure}

Given that there are 5 redshift nodes trained independently, the upper panel of Figure~\ref{fig:combine_2} illustrates the precision of each redshift node, with the root-mean-square relative error (RMSRE) computed for 32 test samples. The RMSRE is articulated as:
\begin{equation}
	\operatorname{RMSRE}(k)=\sqrt{\frac{1}{32} \sum_{i=1}^{32}\left(\frac{P_{\mathrm{emu}}\left(k ; \theta_i\right)}{P_{\mathrm{sim}}\left(k ; \theta_i\right)}-1\right)^2}
\end{equation}
Here, $P_{\mathrm{emu}}\left(k ; \theta_i\right)$ represents the power spectra prediction generated by our emulator, whereas $P_{\mathrm{sim}}\left(k ; \theta_i\right)$ denotes the power spectra derived from simulation data.

The RMSRE functions as a reliable metric for evaluating prediction accuracy. Across 5 nodes, our emulator demonstrates consistent precision. With the exception of significant errors observed at larger scales, the errors produced by our emulator generally remain below 5\%.

\subsection{Comparisons with other methods}\label{sec:compare}
As discussed in previous sections, alternative strategies may exist for constructing our emulator. In this section, we present a comparative analysis of the accuracy of methods that validate our chosen approach as optimal for our emulation scenario, thereby highlighting the importance of incorporating PCA and halofit computations.

To comprehensively gauge the accuracy of the various methods, we computed the RMSREs for the 32 test samples. As illustrated in the lower panel of Figure~\ref{fig:combine_2}, we contrasted three emulation methods: "PCA + HALOFIT," "No HALOFIT," "No PCA," with our "PCA + HALOFIT" method demonstrating superior performance on the test samples.

The "No HALOFIT" method involves using linear power spectra as a reference for generating boost factors instead of employing those computed with halofit. The prediction errors stemming from this approach are notably larger due to the challenge of accurately predicting non-linear BAO features using linear methods. This issue can be effectively addressed by incorporating Mead's halofit method.

Conversely, the "No PCA" method entails training our ANN model directly to predict hundreds of $k$ nodes rather than forecasting 24 PCA coefficients. While both methods yield comparable results with identical training strategies, leveraging PCA results in power spectra with reduced noise, thereby simplifying our training process.

\subsection{Validity of interpolation between redshift nodes}\label{sec:interp}
In this section, we demonstrate the effectiveness of employing the cubic interpolation method for generating predictions between the specified redshift nodes in simulations.

Our emulator is capable of directly predicting power spectra at redshift nodes $z=0.0, 0.5, 1.0, 2.0, 3.0$. To validate our interpolation technique, we calculate the power spectra at $z=0.5$ using interpolation from other redshift nodes and compare these results with those directly derived from simulations. The accuracy of cubic interpolations is depicted in the right panel of Figure~\ref{fig:combine_1}, with relative errors predominantly below 5\%, thereby confirming the efficacy of utilizing cubic interpolations.

\subsection{Example of application}
One of the primary objectives of our emulator is to effectively constrain $f(R)$ cosmology parameters using observational data through Markov chain Monte Carlo (MCMC) methods \citep{1970Bimka..57...97H}. MCMC is a statistical technique employed for sampling from intricate probability distributions, particularly well-suited for high-dimensional parameter spaces and complex posterior distributions. Through the construction of a Markov chain, MCMC methods navigate the parameter space, generating samples based on the characteristics of the target distribution. This process facilitates exploration of parameter space and estimation of posterior distributions, offering a robust framework for Bayesian inference, parameter estimation, and model comparison.

To assess the likelihood of observing the data given specific values of the model parameters, the likelihood function assumes a pivotal role by serving as the foundation for computing the posterior distribution according to Bayes' theorem. This likelihood function can be mathematically represented as:
\begin{equation}
\mathcal{L}(\theta \mid D)=\exp \left(-\frac{1}{2}(\mathbf{O}(D)-\mathbf{M}(\theta))^T \Sigma^{-1}(\mathbf{O}(D)-\mathbf{M}(\theta))\right)
\end{equation}
Here, $\theta$ denotes the model parameters, $D$ signifies the observed data, $\mathbf{O}(D)$ represents a column vector of observables derived from the data, $\mathbf{M}(\theta)$ represents a column vector of model predictions $M_i(\theta)$, and $\Sigma$ denotes the covariance matrix of the errors.

In Bayesian inference, the likelihood function is integrated with a prior distribution over the parameters to calculate the posterior distribution using Bayes' theorem, expressed as:
\begin{equation}
P(\theta \mid D) \propto \mathcal{L}(\theta \mid D) \times \pi(\theta)
\end{equation}
where $\pi(\theta)$ denotes the prior distribution over the parameters. This formula indicates that the posterior distribution of the parameters given the data is proportional to the product of the likelihood function and the prior distribution.

To demonstrate the parameter constraining capabilities of our emulator at small scales, we employ power spectra data at scales where $k > 10^{-1} h \mathrm{Mpc}^{-1}$ obtained from Quijote-MG simulations at different redshift nodes, specifically $z=0.0, 0.5, 1.0, 2.0, 3.0$. Each redshift is considered as an independent observation of the power spectrum, implying that the power spectra at these five redshift nodes are assumed to be uncorrelated. The covariance matrix, essential for our analysis, can be computed using the formula introduced by \citet{1999ApJ...527....1S} as:
\begin{equation}
\Sigma\left(k_1, k_2\right)=\left[\frac{2}{N_{k_1}}+\sigma_{\text{sys}}^2\right] P^2\left(k_1\right) \delta_{k_1 k_2}
\end{equation}
Here, $N_{k_i}=4 \pi k_i^2 \delta k_i/ V_f$ represents the number of independent Gaussian variables in the bin centered on $k_i$ with a width of $\delta k_i$, and $V_f=(2\pi)^3/V$ denotes the fundamental volume of $k$ space in a simulation with volume $V$. For the term $\sigma_{\text{sys}}$, we consider a value of $0.05$ for our small-scale predictions. We employ the Python package \texttt{emcee} \citep{2013PASP..125..306F} for conducting the MCMC analysis.

\begin{figure*}
    \centering
	\includegraphics[width=1.7\columnwidth]{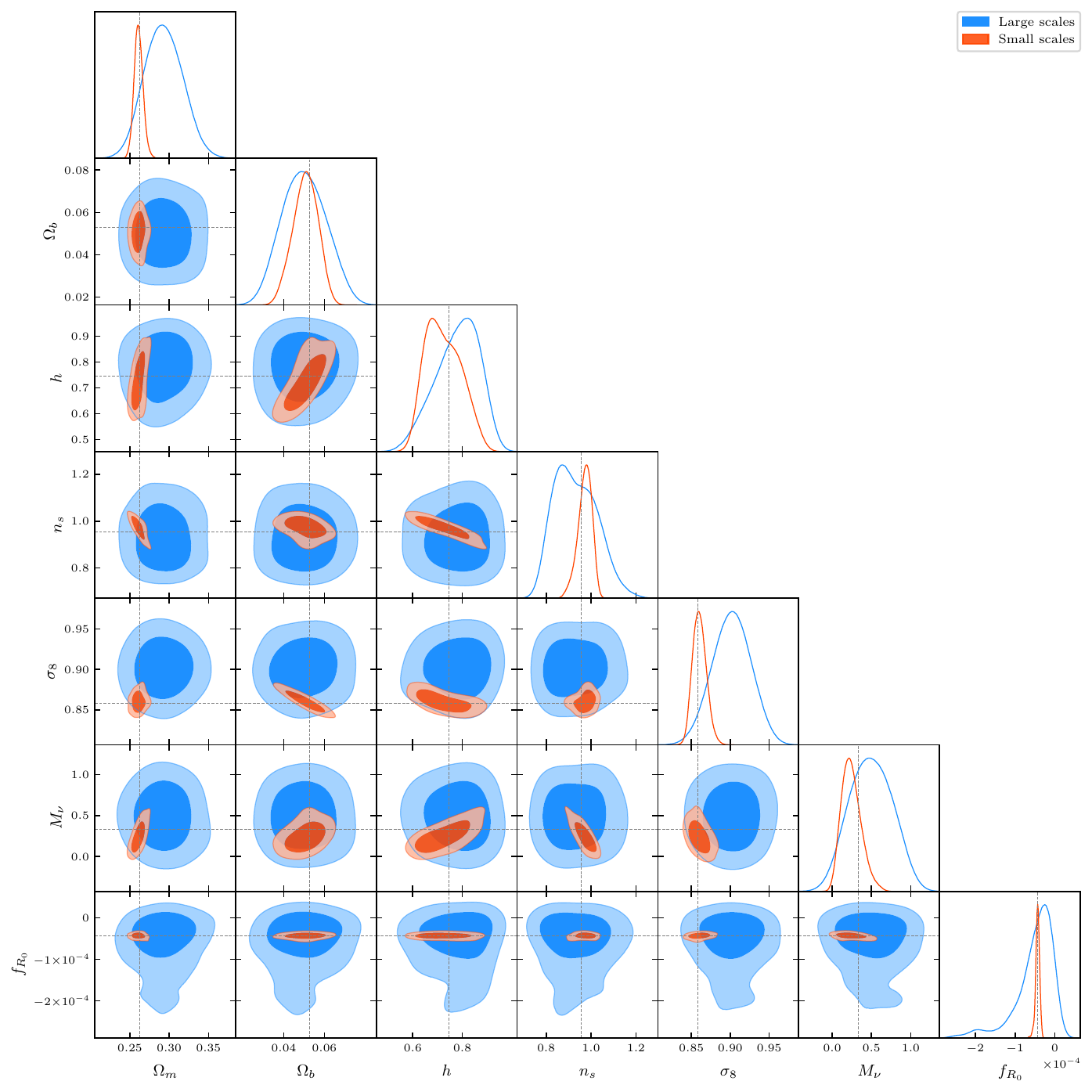}
    \caption{The contour plots illustrate the MCMC results in our study. The blue contours represent constraints derived from large-scale data where $k < 10^{-1} h \mathrm{Mpc}^{-1}$, whereas the red contours depict constraints obtained from small-scale data where $k > 10^{-1} h \mathrm{Mpc}^{-1}$. These findings demonstrate significantly enhanced and more robust constraints on all seven parameters.}
    \label{fig:mcmc}
\end{figure*}

Our MCMC analysis, as illustrated in Figure~\ref{fig:mcmc}, presents the outcomes derived from this methodology. The authentic values of the cosmological parameters are denoted by dashed lines for reference, while the posterior probabilities of the parameters are represented by red contours. The 95\% confidence intervals of the parameters are detailed in the third column of Table~\ref{tab:mcmc}. The slight disparities observed between the true values and the MCMC outcomes can be ascribed to the inherent parameter degeneracy existing in diverse cosmologies. This degeneracy suggests that the power spectra of two sets of parameters may exhibit similar characteristics, posing challenges in accurately constraining the parameters.

One strategy to alleviate parameter degeneracy involves integrating joint constraints from a variety of observational datasets. By amalgamating data from different cosmological probes such as Cosmic Microwave Background (CMB) observations \citep{2020A&A...641A...6P}, Baryon Acoustic Oscillations (BAO), and Type Ia Supernovae (SN) \citep{2018ApJ...859..101S}, we can impose more precise restrictions on cosmological parameters. Joint constraints provide a potent mechanism to resolve the degeneracy and deduce the genuine cosmological parameters governing the evolution of the universe more accurately.

In a bid to underscore the advantages of utilizing small-scale data, supplementary analyses were conducted. Specifically, attention was directed towards scales where $k < 10^{-1} h \mathrm{Mpc}^{-1}$ with the systematic error fixed at $\sigma_{\text{sys}} = 0.1$. MCMC computations were carried out to derive constraints using data at these scales. The findings, depicted in Figure~\ref{fig:mcmc} (blue contours) and Table~\ref{tab:mcmc} (the fourth column), reveal that the efficacy of constraints is significantly diminished in comparison to the utilization of small-scale data. This emphasizes the significance of leveraging small-scale data for more robust parameter estimations.

\begin{table}
	\centering
	\caption{The table presents the true values of parameters alongside the constraint results. The first column lists the names of the parameters, the second column displays their respective true values based on the selected sample. The third column showcases the 95\% confidence limits determined through our emulator and small-scale data from simulations. In contrast, the fourth column represents the 95\% confidence limits derived from large-scale data, specifically where $k < 10^{-1} h \mathrm{Mpc}^{-1}$. Notably, the constraints obtained from small-scale data outperform those from large-scale data.}
	\label{tab:mcmc}
	\begin{tabular} { c c c c }
		
		Parameter & True &  Small-scale & Large-scale \\ 
		\hline
			{\boldmath$\Omega_m$} & 0.263 & $0.261^{+0.012}_{-0.011}   $& $0.293^{+0.045}_{-0.042}   $\\
			
			{\boldmath$\Omega_b$} &0.053 & $0.051^{+0.012}_{-0.012}   $& $0.050^{+0.018}_{-0.018}   $\\
			
			{\boldmath$h$} & 0.746 & $0.72^{+0.14}_{-0.12}      $& $0.78^{+0.13}_{-0.17}      $\\
			
			{\boldmath$n_s$}&0.955 & $0.973^{+0.057}_{-0.066}   $& $0.93^{+0.17}_{-0.14}      $\\
			
			{\boldmath$\sigma_8$} & 0.858 & $0.860^{+0.018}_{-0.017}   $& $0.902^{+0.044}_{-0.046}   $\\
			
			{\boldmath$M_{\nu}$} &  0.332 & $0.24^{+0.26}_{-0.23}      $& $0.50^{+0.48}_{-0.45}      $\\
			
			{\boldmath$10^5f_{R_0}$} &$-4.24$ & $-4.3^{+1.0}_{-1.0}$& $-4.9^{+5.7}_{-10}$\\
			\hline
	\end{tabular}
\end{table}

\section{Conclusions}\label{sec:conclusions}
In this study, we have introduced a novel public emulator named \texttt{FREmu}, specifically crafted to deliver rapid and precise forecasts of non-linear matter power spectra within the HS $f(R)$ gravity model incorporating massive neutrinos. Our primary objective was to address the pressing demand for accurate theoretical predictions accounting for non-linear effects, crucial for the rigorous constraints of $f(R)$ gravity utilizing data from Stage-IV surveys.

The HS $f(R)$ gravity model, a departure from general relativity, emerges as a noteworthy alternative to the conventional $\Lambda$CDM framework. We have elucidated the impact of modifications in $f(R)$ gravity on the formation of cosmic structures and the gravitational potential, underscoring its pertinence in observational examinations.

To develop our emulator, we harnessed data from the Quijote-MG simulation suite, encompassing over 2,000 simulations varying in cosmological parameters. Our emphasis was on predicting power spectrum boosts rather than the power spectra per se, thereby substantially mitigating the computational load. Leveraging PCA to reduce the dimensionality of the target space, we subsequently employed an ANN to efficiently map parameters to power spectrum boosts.

Our precision assessments have revealed that the emulator has achieved accuracy exceeding 95\% in the majority of scenarios, furnishing dependable predictions across a broad spectrum of cosmological parameters and scales. We have exemplified the application of the emulator in parameter constraints through MCMC techniques. The outcomes have demonstrated a solid concordance between the genuine parameter values and those derived from the MCMC analysis, thereby validating the effectiveness of our methodology.

Moreover, we have juxtaposed the efficacy of parameter constraints utilizing small-scale data against large-scale data, showcasing the benefits of leveraging small-scale information in refining parameter constraints.

In summary, \texttt{FREmu} represents a valuable instrument for investigating $f(R)$ gravity models and their implications in cosmology. Its capability to furnish swift and accurate forecasts of non-linear matter power spectra facilitates efficient parameter constraints, thereby facilitating the exploration of modified gravity theories in the non-linear domain. Future endeavors could encompass expanding the emulator to encompass other observables like bispectra and integrating additional observational datasets for comprehensive analyses.

\section*{Acknowledgements}
We thank Francisco Villaescusa and Marco Baldi for providing permission to use the Quijote-MG data and giving us helpful comments. This work is supported by the National Natural Science Foundation of China under grant Nos. U1931202 and
12021003, the National Key R\&D Program of China No.
2020YFC2201603, and the Fundamental Research Funds
for the Central Universities.

\section*{Data Availability}
The emulator presented in this work and its source code is available on GitHub\footnote{\texttt{FREmu} codebase: \url{https://github.com/AstroBai/FREmu}.} under a MIT License and version 1.0 is archived in Zenodo \citep{jiachen2024}. The data generated during our training process can be shared as long as the request is reasonable.


\bibliography{sample631}{}
\bibliographystyle{aasjournal}



\end{document}